\documentclass{article}  
\usepackage{graphicx}
\usepackage{epstopdf}
\title{Muon decay and neutrino masses}  
\author{M.M. Giannini\\
Dipartimento di Fisica dell'Universit\`a di Genova\\
and INFN, Sezione di Genova, Genova, Italy\\
giannini@ge.infn.it
}

\date{}
\begin{document}             
%\linenumbers

\maketitle       
\tableofcontents  

\begin{abstract}
The calculation of the $\mu$ decay is performed in V-A theory taking into account that, in agreement with the observation of the oscillations, the neutrinos must be massive. Provided that the masses of the electron and muon neutrinos  are not higher than the known upper values, the result is a possible modification of the Fermi constant, no effect at all on the polarization of the outgoing electron and possibly an endpoint effect on the electron energy and angle distributions. 

\end{abstract}

\section{Introduction}      

The neutrino oscillation is a well established phenomenon (for a review see \cite{pdg22}), which implies that neutrinos are massive, whereas in the Standard Model they are massless. Presently there is no direct measure  of the neutrino masses and the only information is given by upper limits obtained from the beta decay \cite{aker}, the $\pi^+ decay$ \cite{assam} and the $\tau$ decay \cite{bar}, for the $e$-, $\mu$- and $\tau$-types, respectively. 

In the beta decay the non zero neutrino mass induces a modification on the end point $E_m$ in the electron energy spectrum 
and measuring it one can obtain, as in the KATRIN experiment \cite{aker}, an evaluation of the upper limit of the electron neutrino mass. A similar effect can be expected also for the end point $E_m$ in the muon decay.

In this paper  the muon decay is calculated in  the standard V-A approach taking into account non zero masses for both electron and muon neutrinos in order to put in evidence possible modifications. To this end, various observables are calculated, as the $\mu$ mean lifetime, the electron helicity and the energy and angle distributions of the outgoing electron.

The calculations are performed using neutrino masses within  the present upper values, that is
$m_{\bar{\nu}_e} \leq 0.8 eV$ \cite{aker}and $m_{\nu_\mu} \leq 0.19~ MeV$ \cite{assam}.

\section{The decay rate}

Let us for simplicity concentrate on the decay of the negative muon, that is on the process
\begin{equation}
\mu^- \rightarrow e^- + \bar{\nu}_e + \nu_\mu
\end{equation}

Of course, by applying the charge conjugation, the results obtained in what follows are valid also for the decay
\begin{equation}
\mu^+ \rightarrow e^+ + \nu_e + \bar{\nu}_\mu
\end{equation}

The transition amplitude  is
\begin{equation}
M= \frac{G_F}{\sqrt{2}}[\bar{u}_e(p,s) \gamma^\alpha(1-\gamma_5)\nu_e(q_1,s_1)] [\bar{\nu}_\mu (q_2,s_2) \gamma^\beta(1-\gamma_5)u_\mu(k,t)]
\end{equation}
where $G_F$ is the Fermi constant, ${u}_e (u_\mu)$ is the electron (muon) spinor with energy-momentum and spin p,s (k,t), while $q_1$ and $q_2$ are the fourmomenta of the electron and muon neutrino, respectively, and $s_1, s_2$ are their spin components.

The decay rate is
\begin{equation}
d\Gamma = \frac{1}{(2\pi)^5}\frac{1}{2 k_0}\delta(q-q_1-q_2)|M|^2 \frac{d^3p}{2 p_0} \frac{d^3q_1}{2 q_{10}}\frac{d^3q_2}{2 q_{20}}
\end{equation}
with $q=k-p$.

The transition strength, after summation over the two neutrino spins, is given by
\begin{equation}
\Sigma_{s_1 s_2}|M|^2 =32 G_F^2 (( k - m_\mu t) q_1) ((p - m_e s) q_2)
\end{equation}

The dependence on the neutrino masses becomes evident after having performed the integration  over the neutrino trimomenta. 
In fact, defining the decay rate $d\Gamma_\nu$, summed and integrated over all neutrino trimomenta, we have
\begin{equation}
d\Gamma_\nu = \frac{1}{(2\pi)^4} \frac{G_F^2}{6}\frac{1}{k_0 p_0}[\alpha(q^2) (k - m_\mu t)(p - m_e s) + 2 \beta(q^2)  (q (k - m_\mu t) )(q(p - m_e s))]d^3p
\end{equation} 
where 
\begin{equation}
\alpha(q^2)  =  (q^2 - 2 \Sigma + \frac{\Delta^2}{q^2})g(q^2)~~~~~~~~\beta(q^2) =  (1 + \frac{\Sigma} {q^2} - 2 \frac{\Delta^2}{q^4})g(q^2)
\end{equation}

The quantities $\Sigma, \Delta$ and $g(q)$ depend explicitly on the neutrino masses
\begin{equation}
\Sigma  =  m_{\bar{\nu}_e}^ 2 + m_{\nu_\mu}^2, ~~~~~~~\Delta  =  m_{\nu_\mu}^2- m_{\bar{\nu}_e}^2
\label{ds}
\end{equation}\begin{equation}
g(q^2)=\frac{\sqrt{(q^2-\Sigma)^2-4m_{\nu_e}^2m_{\nu_\mu}^2}}{q^2}
\end{equation}

In the expression of $g(q^2)$, the term with the neutrino masses can be neglected. In fact, the quantity
\begin{equation}
g(q^2)\sim\frac{\sqrt{(q^2-\Sigma)^2}}{q^2}=1-\frac{\Sigma}{q^2}
\end{equation}
is a very good approximation of $g(q^2)$, as it can be numerically verified using the upper limits of the neutrino masses.

From now on we shall perform the calculations in the muon rest frame, where
\begin{equation}
k=(m_\mu, 0), ~~~~ t=(0,\vec{t}), ~~~~p =(E,\vec{p})~~~~s=(\frac{\vec{p} \cdot \vec{n}}{m_e}, \vec{n}+\frac{(\vec{p} \cdot \vec{n})\vec{p}}{m_e(E+m_e)})
\end{equation}
here $m_e$ ($m_\mu$) is the electron (muon) mass and $\vec{n}$ ($\vec{t}$) the corresponding spin vector in the muon rest frame.
In particular we obtain
\begin{equation}
q^2 = 2 m_\mu (W-E)
\end{equation}
where
\begin{equation}
W = \frac{m_\mu^2+m_e^2}{2 m_\mu} =  \frac{m_\mu}{2} (1+x),  ~~~~~~x=\frac{m_e^2}{m_\mu^2}
\end{equation}
 The maximum energy of the electron is
\begin{equation}
E_m = W - \frac{(m_{\bar{\nu}_e}+m_{\nu_\mu})^2} {2 m_\mu}
\label{em}
\end{equation}
and depends on the neutrino masses.

Defining the quantity
\begin{equation}
S = \frac{16}{Em_\mu^6}[\alpha(q^2) (k - m_\mu t)(p - m_e s) + 2 \beta(q^2)  (q (k - m_\mu t) )(q(p - m_e s))]
\label{fac}
\end{equation}
the decay strength can now be written as
\begin{equation}
d\Gamma_\nu =  \frac{1}{(2\pi)^4} m_\mu^5\frac{G_F^2}{96} S d^3p
\label{dgamma}
\end{equation}
and $S$ can be expanded in terms of the various polarization vectors as follows
\begin{equation}
S = S_u + \vec{p} \cdot \vec{t} S_t + \vec{p} \cdot \vec{n} S_n + \vec{n} \cdot \vec{t} S_{nt} + (\vec{p} \cdot \vec{n}) (\vec{p} \cdot \vec{t}) S_{pnt}
\label{dec}
\end{equation} 
 where
\begin{equation}
S_u=\frac{32}{m_\mu^4}[ 3W -2E -\frac{m_e^2}{E}-3\frac{\Sigma} {2m_\mu}+\frac{\Delta^2}{4m_\mu^2}\frac{E-3W+2\frac{m_e^2}{E}}{(W-E)^2}+\frac{\Sigma^2}{4m_\mu^2}\frac{\frac{m_e^2}{E}-E}{(W-E)^2}+
\label{Su}
\end{equation}
\begin{equation}
+\frac{\Sigma \Delta^2}{ 8m_\mu^3}\frac{3W-E-2\frac{m_e^2}{E}}{(W-E)^3}]
\end{equation}
\begin{equation}
S_t=\frac{32}{Em_\mu^4}[ 2W-2E -m_\mu-3\frac{\Sigma} {2m_\mu} +  \frac{\Delta^2}{4m_\mu^2} \frac{5W+E-2m_\mu}{(W-E)^2}+\frac{\Sigma^2}{4m_\mu^2}\frac{m_\mu-E}{(W-E)^2}+
\label{St}
\end{equation}
\begin{equation}
+\frac{\Sigma \Delta^2}{ 8m_\mu^3}\frac{3W -E -2m_\mu}{(W-E)^3}]
\end{equation}

\begin{equation}
S_n=-\frac{32}{Em_\mu^4}[ W-2E+ m_\mu-3\frac{\Sigma}{2m_\mu}+\frac{\Delta^2}{4m_\mu^2}\frac{W + E-2m_\mu}{(W-E)^2}+
\end{equation}
\begin{equation}
+ \frac{\Sigma^2}{4m_\mu^2}\frac{2W-E-m_\mu}{(W-E)^2}+\frac{\Sigma \Delta^2}{ 8m_\mu^3}\frac{-W -E +2m_\mu}{(W-E)^3}]
\label{Sn}
\end{equation}
\begin{equation}
S_{nt}=-\frac{32}{Em_\mu^4}m_e[W-E -3 \frac{\Sigma}{2 m_\mu} + \frac{\Delta^2}{4m_\mu^2}\frac{1}{W-E}+2\frac{\Sigma^2}{4m_\mu^2} \frac{1}{W-E}-\frac{\Sigma \Delta^2}{ 8m_\mu^3}\frac{1}{(W-E)^2}] 
\label{Snt}
\end{equation}
\begin{equation}
S_{pnt}=\frac{32}{Em_\mu^4}[\frac{-W+2E+m_e}{E+m_e} +\frac{\Sigma} {2m_\mu} \frac{3}{E+m_e} -\frac{\Delta^2}{4m_\mu^2}\frac{W+E-m_e}{(W-E)^2(E+m_e)}+
\end{equation}
\begin{equation}
-\frac{\Sigma^2}{4m_\mu^2} \frac{W+E-m_e}{(W-E)^2(E+m_e)}+\frac{\Sigma \Delta^2}{ 8m_\mu^3}\frac{W+E-m_e}{(W-E)^3(E+m_e)}]
\label{Spnt}
\end{equation}

The terms in $\Sigma$ and $\Delta^2$ arise because of the non vanishing neutrino masses. In the expressions of the $S$-quantities there are denominators of the type $(W-E)^{-n}$, with $n=1,2,3$, which cannot diverge thanks to Eq.(\ref{em}), but can be very large. We recall that they are considered keeping the neutrino masses within their present upper values, namely $m_{\bar{\nu}_e}\leq 0.8~ 10^{-6} ~MeV$ and $m_{\nu_\mu} \leq 0.19~ MeV$.

The decay rate of Eq. (\ref{dgamma}) can now be used to calculate various experimental quantities of interest and test within which range a non zero neutrino mass is compatible with the experimental uncertainties.
In the following we shall calculate the muon mean life, the electron helicity and the electron energy and angular distributions.

\section{The muon mean lifetime}

In order to calculate the mean lifetime, one has to average over the muon polarization, to sum over the electron one and integrate over the full range of momentum p or energy E.

The total width is then
\begin{equation}
\Gamma = \frac{m_\mu^5}{\pi^3}\frac{G_F^2}{192}\int_{m_e}^{E_m} dE E p S
\label{gam}
\end{equation}

After having calculated the various integrals, the width is given by
\begin{equation}
\Gamma = \frac{m_\mu^5}{\pi^3}\frac{G_F^2}{192}(Q_1 + \frac{\Sigma}{2 m_\mu^2} Q_2+ \frac{\Delta^2}{4 m_\mu^4} Q_3+\frac{\Sigma^2}{4m_\mu^4} Q_4+\frac{\Sigma \Delta^2}{ 8m_\mu^6}Q_5)
\end{equation}

The $Q_i$ quantities are given by the integrals
\begin{equation}
Q_1 =\frac{32}{m_\mu^4} \int_{m_e}^{E_m}(3 WE  -2 E^2- m_e^2 )pdE
\end{equation}
\begin{equation}
Q_2=-3\frac{32}{m_\mu^3} \int_{m_e}^{E_m}pEdE
\end{equation}
\begin{equation}
Q_3=\frac{32}{m_\mu^2} \int_{m_e}^{E_m}(\frac{E^2-3WE+2m_e^2}{(W-E)^2)} pdE
\end{equation}
\begin{equation}
Q_4=\frac{32}{m_\mu^2} \int_{m_e}^{E_m}\frac{m_e^2-E^2}{(W-E)^2}pdE
\end{equation}
\begin{equation}
Q_5=\frac{32}{m_\mu} \int_{m_e}^{E_m} \frac{3WE-E^2-2m_e^2}{(W-E)^3}pdE
\end{equation}

The terms in $\Sigma$ and $\Delta$ show an explicit dependence on the neutrino masses, but there is also an intrinsic one. In fact, in the case on massless neutrinos, the width $\Gamma_0$ is given by
\begin{equation}
\Gamma_0 = \frac{m_\mu^5}{\pi^3}\frac{G_F^2}{192}Q_0
\end{equation}
with
\begin{equation}
Q_0= 1-8x+8x^3-x^4-12x^2 lnx, ~~~~~~x=(m_e/m_\mu)^2
\end{equation}
and $Q_1$ is different from $Q_0$  because the integration  limit depends on the neutrino masses.

We can then write
\begin{equation}
\Gamma = \frac{m_\mu^5}{\pi^3}\frac{G_F^2}{192}(Q_0 + \Delta Q_\nu)
\end{equation}
where
\begin{equation}
\Delta Q_\nu = Q_1-Q_0 +\frac{\Sigma}{2 m_\mu^2} Q_2+ \frac{\Delta^2}{4 m_\mu^4} Q_3+\frac{\Sigma^2}{4m_\mu^4} Q_4+\frac{\Sigma \Delta^2}{ 8m_\mu^6}Q_5
\end{equation}

\begin{figure}[ht]
\centering
\includegraphics[width=4.5in]{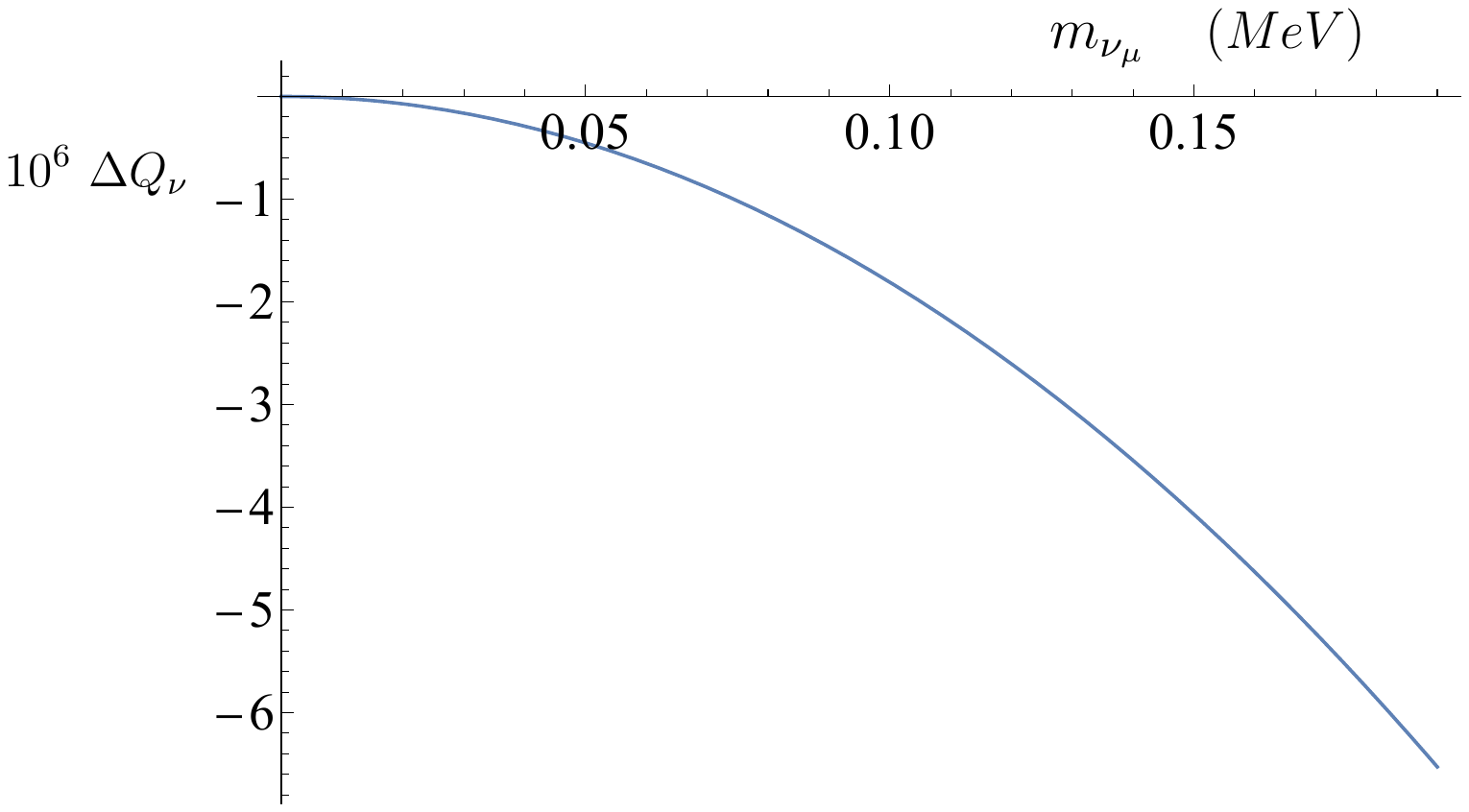}
\caption{Plot of the neutrino correction $\Delta  Q_\nu$  multiplied by $10^5$, as a function of $m_{\nu_\mu} $, with $m_{\bar{\nu}_e}=0.8~  eV$.}
\label{dq}
\end{figure}

Fig. \ref{dq} reports the dependence of $\Delta Q_\nu$ as a function of the muon neutrino mass $m_{\nu_\mu} $ in the range $0-0.19 ~MeV$, showing that the neutrino mass correction is at most of the order of $10^{-6}$. In fact, the maximum value, corresponding to the upper limits of the neutrino masses $m_{\nu_e} \leq 0.8 ~eV , m_{\nu_\mu} \leq 0.19 ~M eV $, is $\Delta Q_\nu=-6.53~ 10^{-6}$.

The numerical contributions of the various terms, corresponding to the maximum values of the neutrino masses,  are easily calculated
\begin{equation}
Q_1-Q_0=-6.47~ 10^{-6}
\end{equation}
\begin{equation}
\frac{\Sigma}{2 m_\mu^2} Q_2=-6.12~ 10^{-8}
\end{equation}
\begin{equation}
 \frac{\Delta^2}{4 m_\mu^4} Q_3=-1.16~ 10^{-9}
\end{equation}
\begin{equation}
\frac{\Sigma^2}{4m_\mu^4} Q_4=-1.16~ 10^{-18}
\end{equation}
\begin{equation}
\frac{\Sigma \Delta^2}{ 8m_\mu^6}Q_5=5.48~ 10^{-12}
\end{equation}
showing that the first two terms are dominant and this is true in the whole energy range. In all these quantities, the main contribution is due to the muon neutrino mass.

From the comparison with the experimental value of the muon mean lifetime $\tau_\mu$, one can obtain the phenomenological value of the Fermi constant $G_F$, but to this end one has to take into account the radiative corrections $\Delta Q_{rad}$:
\begin{equation}
\frac{\hbar}{\tau_\mu} = \frac{m_\mu^5}{\pi^3}\frac{G_F^2}{192}Q_1 (1 + \Delta Q_{rad})
\end{equation}
where, according to \cite{rad}, $\Delta Q_{rad}$ has the value  $-0.00419834(3)$.

One can then write
\begin{equation}
G_F^2 = C\frac{1}{Q_1 (1 + \Delta Q_{rad})}
\end{equation}
having defined
\begin{equation}
C =\frac{\hbar}{\tau_\mu}  \frac{192 \pi^2}{m_\mu^5}
\end{equation}

If one considers also the correction due to the non zero value of the neutrino masses, one has a different Fermi constant $G_{F\nu}$
 \begin{equation}
G_{F\nu}^2 = C\frac{1}{(Q_1+ \Delta  Q_\nu) (1 + \Delta Q_{rad})}
\end{equation}
with a relative variation
\begin{equation}
\Delta_{\Gamma}=|\frac{G_F^2-G_{F\nu}^2}{G_F^2}|= | \frac{ \Delta  Q_\nu}{Q_1 +  \Delta  Q_\nu}| \leq 6.53~ 10^{-6}
\end{equation}

Since the usually quoted Fermi constant, taking into account the radiative corrections, is \cite{rad}
\begin{equation}
G_F = 1.1663788(6) ~10^{-5}~ GeV^{-2}
\end{equation}
with a relative uncertainty of  about $5.1 ~10^{-7}$, it is clear that the neutrino mass correction may affect  the phenomenological value of the Fermi constant. 
Of course the effect vanishes as long as lower values of the muon neutrino mass are considered.

\section{The helicity of the electron}

We consider now the helicity of the final electron, defined as
\begin{equation}
h = \frac{d\Gamma(\uparrow)-d\Gamma(\downarrow)}{d\Gamma(\uparrow)+d\Gamma(\downarrow)}
\end{equation}
$\uparrow$ ($\downarrow$) means that the electron momentum $\vec{p}$ is parallel (antiparallel) to the the spin vector $\vec{n}$.

In the case of unpolarized muon we get
\begin{equation}
h =\frac{(S_u+pS_n)-(S_u-p S_n)}{(S_u+pS_n)+(S_u-p S_n)} = \frac{pS_n}{S_u}
\end{equation}
where $S_u$ and $S_n$ are given by Eqs.( \ref{Su}) and (\ref{Sn}), respectively.

For massless neutrinos, the helicity $h_0$ is simply given by
\begin{equation}
h_0 =-\frac{p}{E}\frac{W-2E+ m_\mu}{3 W - 2 E- \frac{m_e^2}{E} }
\end{equation}
which  starts to be about $-0.99$ already for energies $E\geq 4 MeV$.

The introduction of finite neutrino masses do not alter significantly such value. To see this, we calculate the quantity
\begin{equation}
\frac{h-h_o}{h_0}
\end{equation}

\begin{figure}[ht]
\centering
\includegraphics[width=5.5in]{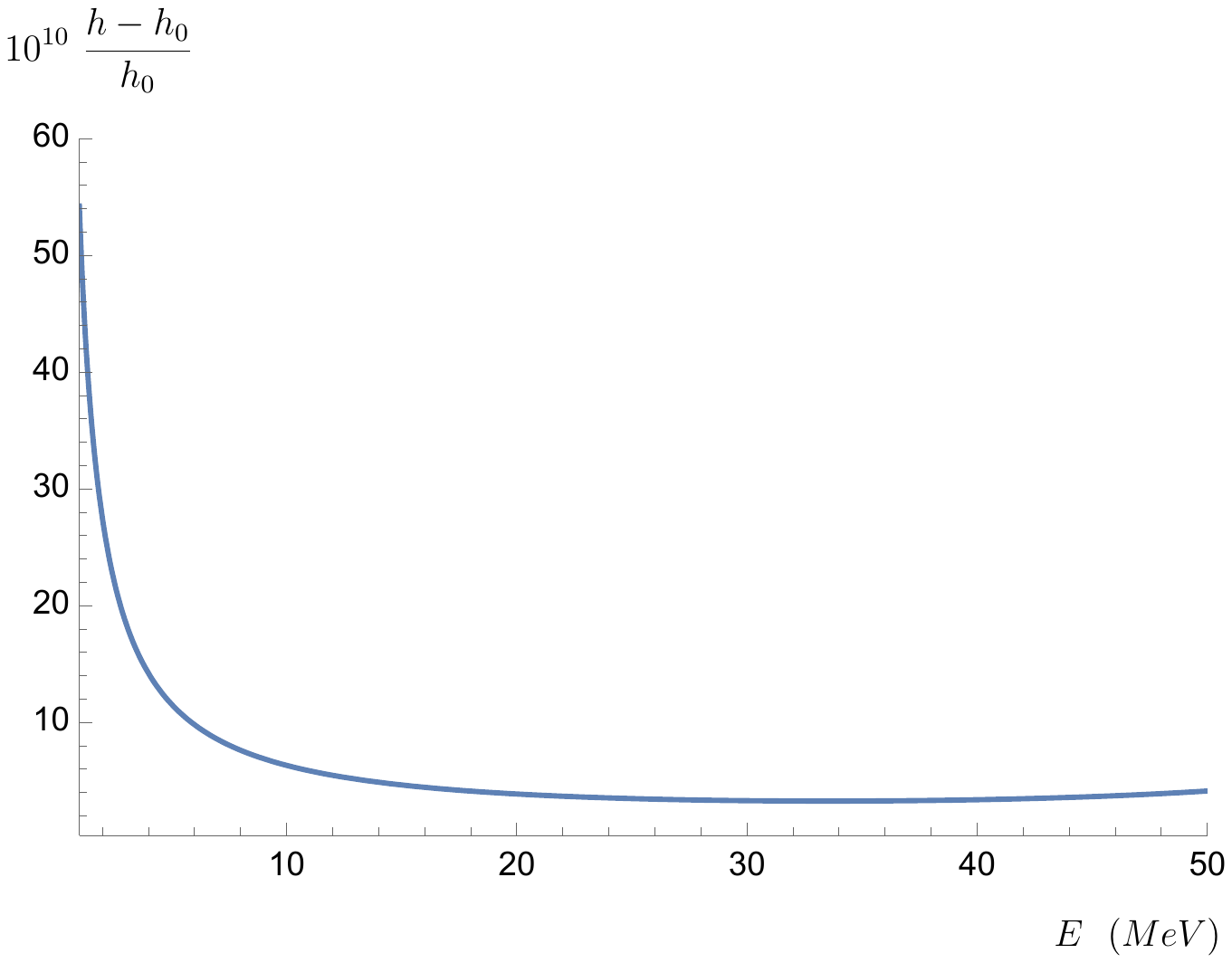}
\caption{Relative variation of the helicity $h$ with respect to the case of massless neutrinos, multiplied by $10^{10}$.}
\label{h}
\end{figure}

As it is shown in Fig. \ref{h}, in presence of a non zero neutrino mass, the relative modification of the electron helicity  is of the oder of $10^{-9}$ for small electron energies and it decreases sharply for increasing energies.  As long as the muon neutrino mass decreases, the relative variation reported in Fig. \ref{h} becomes even smaller.

Since the observed polarization is $1.00 \pm 0.04$ \cite{pdg22}, it is clear  that the introduction of massive neutrinos is perfectly compatible with phenomenology, provided that the $\bar{\nu}_e$  and $\nu_{\mu}$ masses are not larger than the present upper limits.

\section{Energy and angular distribution of the electron}

 Let us consider the case in which both the electron and the muon are polarized. 
 
 The transition strength given by Eq. (\ref{dgamma}) 
\begin{equation}
d\Gamma_\nu =  \frac{1}{(2\pi)^4} m_\mu^5\frac{G_F^2}{96} S d^3p
\end{equation}

Plotting the values of the various  $S$-quantities as  functions of the electron energy $E$, the curves with and without neutrino masses are practically coincident, provided that $E$ is not too near to the end point, which, according to Eq. (\ref{em}), is slightly greater than $52.8 ~MeV$.

In proximity of the end point, that is for $E \geq 52.7 MeV$, the situation is slightly different. 

\begin{figure}[ht]
\centering
\includegraphics[width=5.5in]{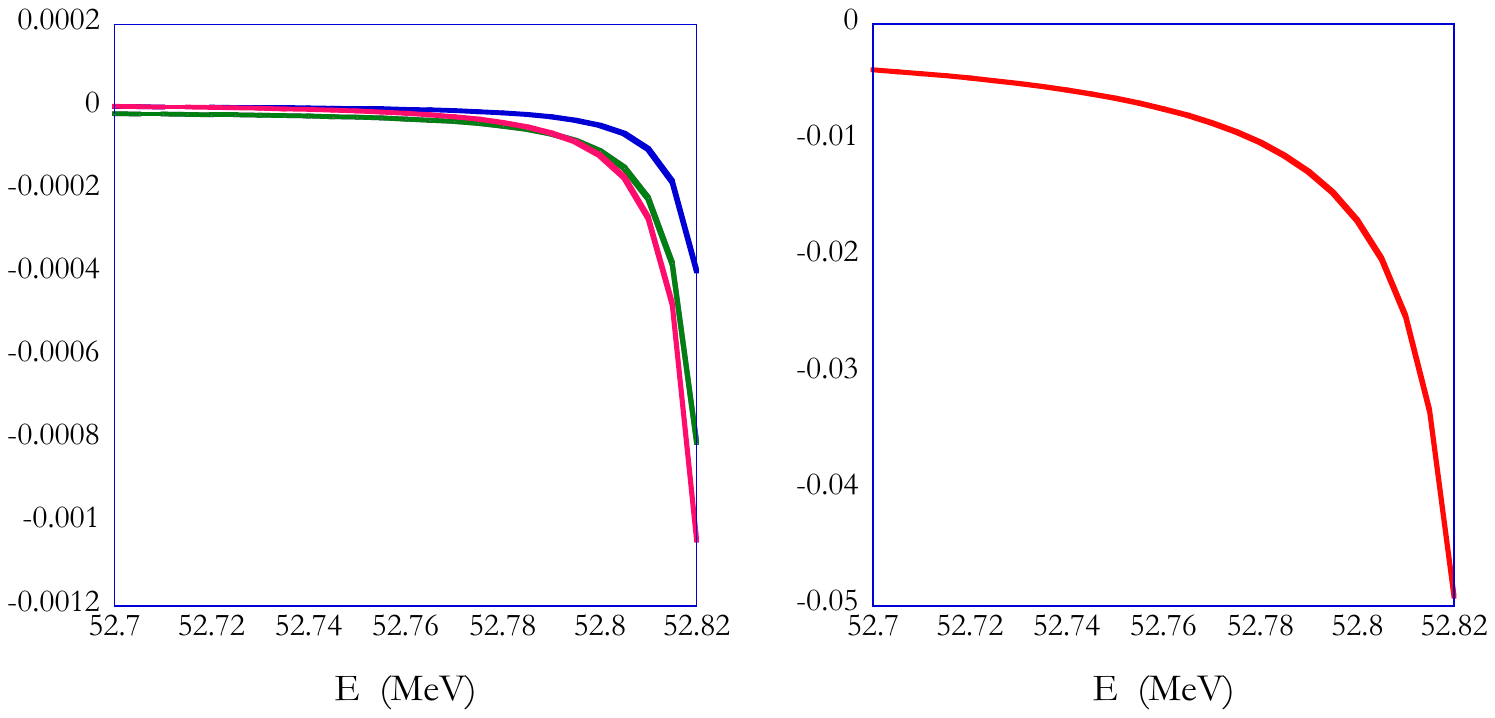}
\caption{The energy dependence of the relative variations $\frac{S-S_0}{S_0}$ for each $S$- function at the end point, $S_0$ being the result without the neutrino contribution. Left:   $S_u$ (green curve), $S_t$ (blue curve), $S_n$ (coincident with $S_u$), $S_{pnt}$ (red curve). Right: $S_{nt}$.}
\label{rel} 
\end{figure}

The relative variations $\frac{S_k-S_{k0}}{S_{k0}}$ are reported in Fig. (\ref{rel}). 
For $k=u, t, n, pnt$ the corresponding results are very similar and very low, being at most of the order of $10^{-4}$. 
In the case of the $S_{nt}$ function the relative difference is of the order of a few percent.
This effect is due to the mass of the muon neutrino, which is assumed to its upper value of $0.19 ~MeV$. Of course, this effect becomes negligible  as long as the muon neutrino mass decreases.

A particular case of interest is provided by the study of the differential decay probability at definite angles, which is given by
\begin{equation}
\frac{d\Gamma(\theta,\phi)}{dcos\theta d\phi dE}=\frac{G_F^2}{(2\pi)^4} \frac{m_\mu^5}{96} S p  E
\end{equation}

Assuming that the muon polarization is parallel to the z-axis and the electron is emitted in the the $x-y$ plane, with its spin parallel ($\uparrow$) or antiparallel ($\downarrow$)  to the muon polarization, we  have
\begin{equation}
\vec{t}=(0,0,1)~ ~ ~ \vec{p}=p(cos\phi, sen\phi,0) ~ ~ ~ \vec{p}\cdot \vec{t} = 0~ ~ ~ \vec{p} \cdot \vec{n} =0 ~ ~ ~\vec{n} \cdot \vec{t} = 1( -1)
\end{equation}

Therefore, integrating over $\phi$

\begin{equation}
\frac{d\Gamma(\uparrow)(\theta)}{dcos\theta dE}=\frac{G_F^2}{(2\pi)^4} \frac{m_\mu^5}{96} S(\uparrow) p  E ~ ~ ~\frac{d\Gamma(\downarrow)(\theta)}{dcos\theta dE}=\frac{G_F^2}{(2\pi)^4} \frac{m_\mu^5}{96} S(\downarrow) p  E
\end{equation}
with
\begin{equation}
S(\uparrow) = S_u+S_{nt} ~ ~ ~~ ~ ~S(\downarrow) = S_u - S_{nt}
\end{equation}

Defining the asymmetry $A$ as
\begin{equation}
A(E) = \frac{\frac{d\Gamma(\uparrow)(\pi/2)}{dcos\theta dE}-\frac{d\Gamma(\downarrow)(\pi/2,}{dcos\theta dE}}{\frac{d\Gamma(\uparrow)(\pi/2)}{dcos\theta dE}+\frac{d\Gamma(\downarrow)(\pi/2,}{dcos\theta dE}}
\end{equation}
we have
\begin{equation}
A(E)  = \frac{S_{nt}}{S_u}
\end{equation}
introducing the explicit expressions of $S_u, S_{nt}$
\begin{equation}
A(E)  = \frac{-\frac{m_e}{E}[W-E -3 \frac{\Sigma}{2 m_\mu} + \frac{\Delta^2}{4m_\mu^2}\frac{1}{W-E}+2\frac{\Sigma^2}{4m_\mu^2} \frac{1}{W-E}-\frac{\Sigma \Delta^2}{ 8m_\mu^3}\frac{1}{(W-E)^2}] }{f_1-3\frac{\Sigma} {2m_\mu}+\frac{\Delta^2}{4m_\mu^2}\frac{f_2}{(W-E)^2}+\frac{\Sigma^2}{4m_\mu^2}\frac{f_3}{(W-E)^2}+\frac{\Sigma \Delta^2}{ 8m_\mu^3}\frac{f_4}{(W-E)^3}}
\end{equation}

One should note that this asymmetry is not affected by any modification of the Fermi constant due to the presence of massive neutrinos.

For massless neutrinos
\begin{equation}
A_0(E)  = -\frac{m_e(W-E) }{3W E -2E^2 -m_e^2}
\end{equation}

The corresponding relative variation is
\begin{equation}
\frac{A(E) -A_0(E) }{A_0(E) }
\end{equation}

\begin{figure}[ht]
\centering
\includegraphics[width=4.5in]{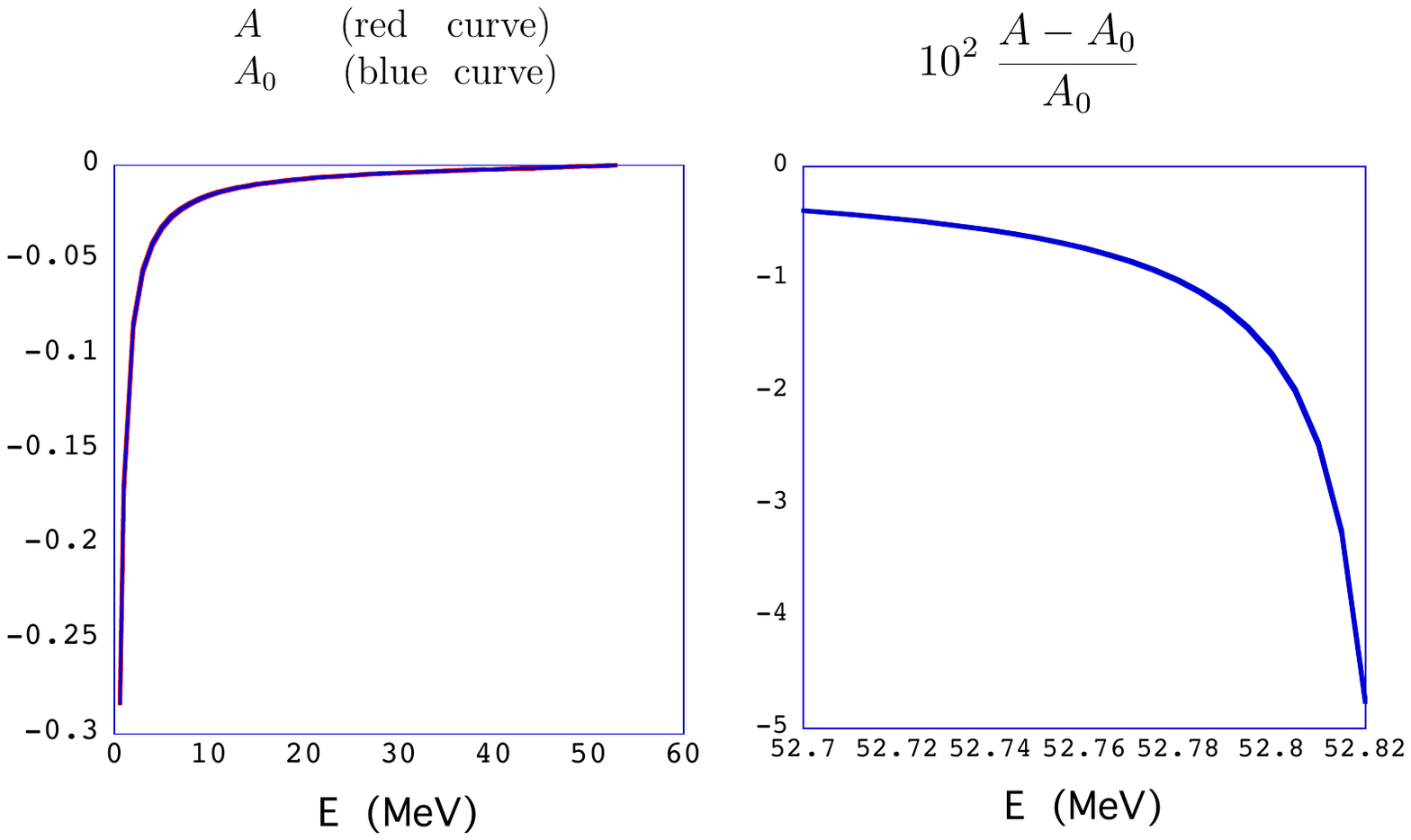}
\caption{Left: the asymmetry as a function of energy. The curves with and without the neutrino mass correction are practically coincident. Rigth: the relative variation of the asymmetry $A$ with respect to the case of massless neutrinos, multiplied by $10^2$ near the end point.}
\label{asym}
\end{figure}

The above asymmetry is determined by the quantity $S_{nt}$, which seems to have an appreciable sensitivity to the presence of massive neutrinos. The Fig. (\ref{asym}) shows that there is a relevant end point effect, of the order of $10^{-2}$.

\section{Conclusions}

Thanks to the results of the Katrin experiment, the electron neutrino mass, if different from zero, is so small that certainly it does not affect the various experimental observables. As for the muon neutrino mass, the situation may be slightly different. If the mass is not far from its upper limit, there is a  influence on the width, with possibly a renormalization of the Fermi constant. The remaining observables, that is electron helicity, energy and angle distribution, are not affected, with the exception of an end point effect near the maximum electron energy $E_m$. 

All these effects are of relative importance and become lower and lower, for decreasing values of the muon neutrino mass.

\end{document}